\documentclass[pre,aps,groupaddress,showpacs,twocolumn]{revtex4-1}
\usepackage{mathrsfs}
\usepackage{epsfig,amsmath,graphicx,amssymb,overpic,float}
\usepackage{epstopdf}
\usepackage{colordvi}
\usepackage{color}
\def\be{\begin{equation}}
\def\ee{\end{equation}}
\def\bee{\begin{eqnarray}}
\def\ene{\end{eqnarray}}
\def\bes{\begin{subequations}}
\def\ees{\end{subequations}}

\begin{document}
\title{\large Nonautonomous discrete rogue waves and interaction in the  generalized \\
Ablowitz-Ladik-Hirota lattice with variable coefficients}
\author{Zhenya Yan$^{1}$}
\author{Dongmei Jiang$^{1,2}$}
\author{W. M. Liu$^{3}$}

\affiliation{\vspace{0.05in} $^1\!$Key Laboratory of Mathematics Mechanization,
Institute of Systems Science, AMSS, Chinese Academy of Sciences,
Beijing 100190, China\\
$^2\!$Department of  Mathematics, Qingdao University
of Technology,  Qingdao 266033, China\\
$^{3}\!$Beijing National Laboratory for Condensed Matter Physics,
Institute of Physics,
 Chinese Academy of Sciences, Beijing
100190, China   }


\begin{abstract}
 We analytically investigate the nonautonomous discrete rogue wave solutions and their interaction in the generalized Ablowitz-Ladik-Hirota lattice with variable coefficients, which possess complicated wave propagations in time and are beyond the usual discrete rogue waves. When the amplitude of the tunnel coupling coefficient between sites decreases, these nonautonomous discrete rogue wave solutions become localized in time
 after they propagate over some certain large critical values. Moreover, we find that the interaction between nonautonomous discrete rogue waves is elastic. In particular, these results can reduce to the usual discrete rogue wave solutions when the gain or loss term is ignored.

\end{abstract}
\pacs{05.45.Yv, 42.65.Tg, 42.65.Wi}

\maketitle


\section{Introduction}

Rogue waves (alias as {\it freak waves, monster waves, killer waves, giant waves} or {\it extreme waves}), as an important physical phenomenon, are localized both
in space and time and depict a unique event that `appears from nowhere and disappears without a trace'~\cite{prg}. Rogue waves (RWs) are also known as `rogons' if they reappear virtually unaffected in size or shape shortly after their interactions~\cite{vrg}. The study of RWs has become a significant subject in many fields since they can signal catastrophic phenomena such as thunderstorms, earthquakes, and hurricanes. RWs have been found in the ocean~\cite{org,org2}, nonlinear optics~\cite{rge1,rge2,rge3}, Bose-Einstein condensates~\cite{mrg}, the atmosphere~\cite{arg}, and even the finance~\cite{frg}. Moreover, some experimental observations have shown that optical RWs do exist and play a positive role in nonlinear optical fibres~\cite{rge1,rge2,rge3}. Such optical RWs differ from oceanic RWs that play a negative role and lead to many accounts of such waves hitting passenger ships, container ships, oil tankers, fishing boats, and offshore and coastal structures, sometimes with catastrophic consequences~\cite{org2}. In particular,
the {\it analytical} RWs have been obtained for the nonlinear Schr\"odinger (NLS) equation~\cite{nls,nls2,prg}, as well as some of their extensions with the varying coefficients~\cite{vrg}, the higher orders~\cite{ehnls}, or the higher dimensions~\cite{hrg}.

The discrete NLS equation~\cite{Dnls} and the Ablowitz-Ladik (AL) lattice~\cite{AL1,AL2}, as two prototypical discretizations of the continuum NLS equation, have been studied extensively in the field of nonlinear science. The former is nonintegrable, but has some interesting applications of physics~\cite{dnls,dnls2,dnls3,dnls4,dnlspr}. The latter is integrable and possesses an infinite number of conservation laws~\cite{AL1,AL2}, as well as has been depicted as an effective lattice to study properties of the intrinsic localized modes~\cite{ALP}. Moreover the nonintegrable discrete NLS equation can also be regarded as a perturbation of the integrable  AL lattice~\cite{NLSP}.
In addition, the Salerno model (SM) has also been presented, interpolating between the nonintegrable discrete NLS equation ($\mu=0$) and the intregable AL lattice ($\epsilon=0$) in the form~\cite{SM,SMb,SMc,SMd,SMe,SMf}
\begin{eqnarray}
\label{GNLS}
i\psi_{n,t}\!+\!(\psi_{n\!+\!1}\!\!+\!\psi_{n\!-\!1})(1\!+\!\mu|\psi_n|^2\!)\!\!+\epsilon |\psi_n|^2\psi_n\!+v\psi_n=\!0, \quad
 \end{eqnarray}
which can be derived on the basis of a variational principle $\delta\mathcal{L}_{\rm SM}/\delta \psi_n^{*}=0$ from the Lagrangian density
\bee
\nonumber \mathcal{L}_{\rm SM}\!=\!\sum_n i(\psi_n^{*}\psi_{n,t}\!-\!\psi_n\psi^{*}_{n,t})\!\!+4\, {\rm Re}\,(\psi_n^{*}\psi_{n+1})\qquad\qquad \\
+\mu(\psi_{n\!+\!1}\!\!+\psi_{n\!-\!1})\psi_n^{*}|\psi_n|^2\!+\epsilon |\psi_n|^4+2v|\psi_n|^2,\quad\,\,
\ene
where $\psi_n\equiv \psi_n(t)$ stands for the complex field amplitude at the $n$th site of the lattice, the parameter $\mu$ is the intersite nonlinearity and corresponds to the nonlinear coupling between nearest neighbors,
$\epsilon $ measures the intrinsic onsite nonlinearity, and $v$ describes the inhomogeneous frequency shift.
 The SM has been applied in biology~\cite{SM}, and Bose-Einstein condensates~\cite{SMf}. Recently, the discrete RWs have also drawn much attention. The discrete NLS equation (i.e., for the case $\mu=0,\, \epsilon =1$, and $v=-2$ in Eq.~(\ref{GNLS})) has numerically been verified to support discrete RWs~\cite{drg}. The SM has also been found to admit discrete RWs from
the viewpoint of statistical analysis~\cite{SMg}. More recently, it has also been shown that exact discrete RWs~\cite{al,al2} can exist in the AL lattice (i.e., for the case $\mu=1,\, \epsilon =0$, and $v=-2$ in Eq.~(\ref{GNLS})) on the basis of the limit cases of their multi-soliton solutions~\cite{hirota}.
However, there is no reports to date about nonautonomous discrete RWs except for the AL lattice and the discrete Hirota equation~\cite{al,al2}.

    In the present paper, we will explore exact nonautonomous discrete RW solutions and interaction of the generalized Ablowitz-Ladik-Hirota (ALH) lattice with variable coefficients given by Eq.~(\ref{ghirota}), i.e., the generalized case of Eq.~(\ref{GNLS}) without the intrinsic onsite nonlinearity, where the tunnel coupling coefficients and the intersite nonlinearity are the time-modulated, complex-valued and real-valued functions, respectively, the inhomogeneous frequency shift are space- and time-modulated, real-valued functions, and the time-dependent gain or loss term is added. To do so, we will make use of the differential-difference symmetry analysis that can connect this equation with variable coefficients with the simpler ones. We show that, for the attractive intersite nonlinearity, the generalized ALH lattice with variable
coefficients can support nonautonomous discrete RWs in terms of the rogue wave solutions of the discrete Hirota equation. Moreover, we also exhibit wave propagations of nonautonomous discrete RW solutions and their interaction for some chosen parameters and functions.

The rest of this paper is organized as follows. In Sec. II, we introduce the generalized ALH lattice with variable coefficients, which contains some special lattice models, such as the AL lattice, the discrete Hirota equation, and the generalized AL lattice. In Sec. III, by analyzing the phase of the complex field amplitude, we systematically
present a similarity transformation reducing the generalized ALH lattice with variable
coefficients to the discrete Hirota equation. In Sec. IV, we determine the self-similar variables and constraints satisfied by the coefficients in Eq.~(\ref{ghirota}). Moreover, we analyze the relations among these functions such that we find that  the intersite nonlinearity (the inhomogeneous frequency shift) is related to  the gain or loss term (the tunnel coupling coefficient). Sec. V mainly discusses nonautonomous discrete RW solutions and their interaction of Eq.~(\ref{ghirota}) for some chosen parameters and functions.  For the given periodic gain or loss term, when the amplitude of the tunnel coupling coefficient between sites decreases, these nonautonomous discrete rogue wave solutions are localized in space and keep the localization longer in time, which differ from the usual discrete rogue waves. Finally, we give some conclusions in Sec. VI.

\section{The generalized Ablowitz-Ladik-Hirota lattice with variable coefficients}

We here address the generalized Ablowitz-Ladik-Hirota (ALH) lattice with variable coefficients modeled by the following lattice
 \bee
\label{ghirota}
 \begin{array}{l}
 i\Psi_{n,t}+\big[\Lambda(t)\Psi_{n+1}+\Lambda^{*}(t)\Psi_{n-1}\big]\big[1+g(t)|\Psi_n|^2\big]  \vspace{0.1in} \cr
 \quad \quad\,\,\, -2v_n(t)\Psi_n+i\gamma(t)\Psi_n=0,
 \end{array}
 \ene
 which can be derived in terms of  a variational principle $\delta\mathcal{L}_{\rm ALH}/\delta \Psi_n^{*}=0$ from the
 following Lagrangian density
\bee
\begin{array}{l}
\mathcal{L}_{\rm ALH}\!=\!\sum_n i(\Psi_n^{*}\Psi_{n,t}\!-\!\Psi_n\Psi^{*}_{n,t})
 + 4\,{\rm Re}\,[\Lambda(t)\Psi_n^{*}\Psi_{n+1}] \quad \vspace{0.1in} \cr
\qquad\qquad +g(t)[\Lambda(t)\Psi_{n+1}\!+\!\Lambda^{*}(t)\Psi_{n-1}]\Psi_n^{*}|\Psi_n|^2 \vspace{0.1in} \cr
\qquad\qquad  -2[2v_n(t)\!-\!i\gamma(t)]|\Psi_n|^2,
\end{array} \ene
where $\Psi_n\equiv \Psi_n(t)$ stands for the complex field amplitude at the $n$th site of the lattice, the complex-valued function $\Lambda(t)$ is the coefficient of tunnel coupling
between sites and can be rewritten as $\Lambda(t)=\alpha(t)+i\beta(t)$ with $\alpha(t)$ and $\beta(t)$ being differentiable, real-valued functions, $g(t)$ stands for the time-modulated intersite nonlinearity, $v_n(t)$ is the space- and time-modulated inhomogeneous frequency shift, and $\gamma(t)$ denotes the time-modulated effective gain or loss term.

In fact, this nonlinear lattice model (\ref{ghirota}) contains many special lattice models, such as the AL lattice for the case $\alpha(t)={\rm const.},\, \beta(t)=v_n(t)=\gamma(t)=0$, and $g(t)={\rm const.}$~\cite{AL1,AL2}, the AL equation with additional term accounding for dissaption for the case $\alpha(t)={\rm const.},\,\beta(t)=v_n(t)=0,\, \gamma(t)={\rm const.}$, and $g(t)={\rm const.}$~\cite{SMb},  the discrete Hirota equation for the case $\alpha(t)={\rm const.},\,\, \beta(t)={\rm const.},\, v_n(t)=\gamma(t)=0$, and $g(t)={\rm const.}$~\cite{hirota}, the generalized AL lattice given by Eq.~(\ref{GNLS}) for the case $\alpha(t)={\rm const.},\, \beta(t)=\gamma(t)=0$, and $g(t)={\rm const.}$~\cite{dgp}, and the discrete modified KdV equation  for the case $\alpha(t)=v_n(t)=\gamma(t)=0,\, \beta(t)={\rm const.}$, and $g(t)={\rm const.}$~\cite{mKdV}.

\section{Differential-difference similarity reductions and constraint equations}

We consider the spatially localized solutions of Eq.~(\ref{ghirota}), i.e., $\lim_{|n|\to \infty}|\Psi_n(t)|=0$.
To this aim, we search for a proper similarity transformation connecting solutions of
Eq.~(\ref{ghirota}) with those of the following discrete Hirota equation with constant coefficients~\cite{hirota}, namely
\begin{eqnarray}
\label{hirota}
 i\Phi_{n,\tau}\!\!+\!\big(\lambda\Phi_{n+1}\!+\!\lambda^{*}\Phi_{n-1}\big)\!\big(\!1\!+\!|\Phi_n|^2\!\big)\!-\!
 2{\rm Re}\,(\lambda)\Phi_n\!=\!0,
\quad
 \end{eqnarray}
 which can be derived in terms of  a variational principle $\delta\mathcal{L}_{\rm H}/\delta \Psi_n^{*}=0$ from the
 following Lagrangian density
\bee
\begin{array}{l}
\mathcal{L}_{\rm H}\!=\!\sum_n i(\Phi_n^{*}\Phi_{n,\tau}\!-\!\Phi_n\Phi^{*}_{n,\tau})
 + 4\,{\rm Re}\,(\lambda\Phi_n^{*}\Phi_{n+1}) \quad\qquad \vspace{0.1in} \cr
 \qquad\quad +(\lambda\Phi_{n+1}\!+\!\lambda^{*}\Phi_{n-1})\Phi_n^{*}|\Phi_n|^2\! -\!4\,{\rm Re}\,(\lambda)|\Phi_n|^2,
\end{array} \ene
where $\Phi_n\equiv \Phi_n(\tau)$ is a complex dynamical variable at the $n$th site of the lattice, $\tau\equiv \tau(t)$ is a real-valued function of time to be determined, and the complex-valued parameter $\lambda$ can be rewritten as $\lambda=a+ib$ with $a$ and $b$ being real-valued parameters. The discrete Hirota model (\ref{hirota}) contains some special physical models, such as the AL lattice for the case $a=1$ and $b=0$~\cite{AL1, AL2} and the discrete mKdV equation for the case $a=0$ and $b=1$~\cite{mKdV}. It has been shown in Ref.~\cite{dis} that the discrete Hirota equation (\ref{hirota}) is in fact an integrable discretization of the three-order NLS equation (also known as the Hirota equation)~\cite{hirotac}
\bee \label{hnls}
iq_t+a(q_{xx}+|q|^2q)-ib(q_{xxx}+6|q|^2q_x)=0,
\ene
which plays an important role in nonlinear optics~\cite{Optics, hnls}.

To show the above-mentioned aim, we here apply the similarity transformation in the form
\begin{eqnarray}
\label{tran}
 \Psi_n(t)=\rho (t)e^{i \varphi_n(t)}\Phi_n[\tau(t)]
 \quad
 \end{eqnarray}
  to Eq.~(\ref{ghirota}), where the function $\rho(t)$ and phase $\varphi_n(t)$ are both real-valued functions of indicated variables to be determined. To conveniently substitute ansatz (\ref{tran}) into Eq.~(\ref{ghirota}) and to further balance the phases in every term in Eq.~(\ref{ghirota}), i.e., $\Psi_{n+1}(t)$, $\Psi_n(t)$, and $\Psi_{n-1}(t)$,  we should firstly know the explicit expression of the phase $\varphi_n(t)$ in transformation (\ref{tran}) in space.

 Here we consider the case that the phase is expressed as a quadratic polynomial in space with coefficients being functions of time in the form $\varphi_n(t)=p_2(t)n^2+p_1(t)n+p_0(t)$
  with $p_{0,1,2}(t)$ being functions of time, which is similar to the phases in the continuous NLS (or GP) equations with variable coefficients~\cite{SA}. Based on the symmetry analysis, we balance the coefficients of these terms $\Psi_{n+1}(t)$, $\Psi_n(t)$, and $\Psi_{n-1}(t)$ such that we find that the phase in transformation (\ref{tran}) should be a first degree polynomial in space with coefficients being functions of time, namely
 \bee
  \label{phase} \varphi_n(t)=p_1(t)n+p_0(t), \ene
  where $p_{0,1}(t)$ are functions of time to be determined.

 Eq.~(\ref{tran}) with the condition (\ref{phase}) allows us to
reduce Eq.~(\ref{ghirota}) to Eq.~(\ref{hirota}), variables in this reduction can be determined from the
requirement for the new complex field amplitude  $\Phi_n(\tau(t))$ to
satisfy Eq.~(\ref{hirota}). Thus, we
substitute transformation (\ref{tran}) into Eq.~(\ref{ghirota}) along with Eq.~(\ref{phase})
and after relatively simple algebra obtain the following system of
ordinary differential equations
\bes \label{sys}\bee
 \label{sys1} &&  \dot{\rho}(t)+\gamma(t)\rho(t)=0,\\
 \label{sys2} &&  a\dot{\tau}(t)-\alpha(t)\cos p_1(t)+\beta(t)\sin p_1(t)=0,\\
\label{sys3}&&  \left[b\beta(t)+a\alpha(t)\right]\sin p_1(t)+\left[a\beta(t)-b\alpha(t)\right] \nonumber\\
 &&\qquad \times\cos p_1(t)=0, \\
\label{sys4} && 2v_n(t)+\dot{p}_1(t)n+\dot{p}_0(t)+2[\beta(t)\sin p_1(t)\nonumber\\
 && \qquad -\alpha(t)\cos p_1(t)]=0, \\
 \label{sys5} && g(t)\rho^2(t)=1,
\ene
\ees
where the dot denotes the derivative with respect to time.

Therefore, if system (\ref{sys}) is consistent, then we have constructed an algorithm generating nonautonomous  solutions of Eq.~(\ref{ghirota}) based on transformation (\ref{tran}) and solutions of Eq.~(\ref{hirota}):

 Firstly, we solve Eqs.~(\ref{sys1})-(\ref{sys3}) to obtain the functions $\rho(t),\ \tau(t)$, and $p_1(t)$ in transformation (\ref{tran}).

 And then we consider
Eqs.~(\ref{sys4}) and (\ref{sys5}) to determine the inhomogeneous frequency shift $v_n(t)$ and the intersite nonlinearity $g(t)$ in Eq.~(\ref{ghirota}) in terms of above-obtained functions $\rho(t),\ \tau(t)$, and $p_1(t)$.

Thus, we have established a similarity transformation (\ref{tran}) connecting solutions of Eq.~(\ref{hirota}) and those of Eq.~(\ref{ghirota}). In particular, we here exhibit our approach in terms of two lowest-order discrete rogue wave solutions of Eq.~(\ref{hirota}) as seeding solutions to find nonautonomous discrete rogue wave solutions of Eq.~(\ref{ghirota}).

\begin{figure*}
\begin{center}
\vspace{0.05in}
{\scalebox{0.65}[0.6]{\includegraphics{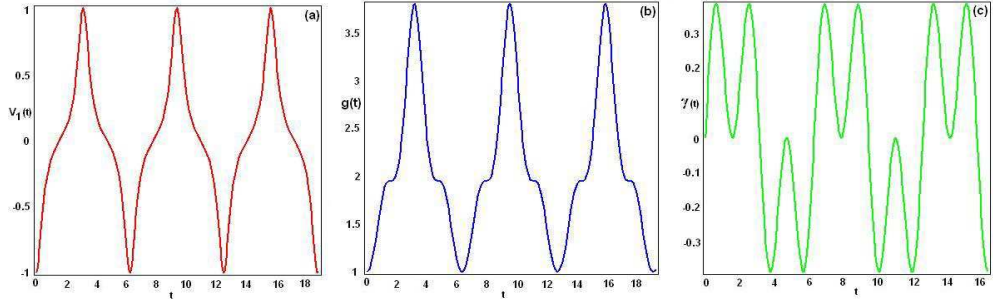}}}
\end{center}
\vspace{-0.15in} \caption{\small (color online). Profiles of the coefficients of the generalized ALH lattice with variable coefficients given by Eq.~(\ref{ghirota}) vs time for the parameters given by Eq.~(\ref{gamm}) with parameters $\gamma_0=c_1=c_2=1$. (a) the coefficient $v_1(t)$ given by Eq.~(\ref{v1}) of the first degree term of the inhomogeneous frequency shift $v_n(t)$ given by Eq.~(\ref{coe1}), (b) nonlinearity $g(t)$ given by
Eq.~(\ref{coe2}), and (c) the gain or loss term $\gamma(t)$.} \label{fig:coe}
\end{figure*}

\section{Determining the similarity transformation and coefficients }

It follows from Eqs.~(\ref{sys1})-(\ref{sys3}) that we can obtain the variables
$\rho(t),\, p_1(t)$ and $\tau(t)$ in transformation (\ref{tran}) in the form
\bes \label{var} \bee
 \label{var1} &&
 \rho(t)=\rho_0\exp\left[-\int^t_0\gamma(s)ds\right], \\
 \label{var2}&&p_1(t)=\tan^{-1}\left[\frac{b\alpha(t)-a\beta(t)}{a\alpha(t)+b\beta(t)}\right],\\
 \label{var3} &&\tau(t)=(a^2\!+b^2)^{-1/2}\int^t_0[\alpha^2(s)+\beta^2(s)]^{1/2}ds,
\ene \ees
where $\rho_0$ is an integration constant. Now it follows from Eqs.~(\ref{sys4}) and (\ref{sys5}) along with Eqs.~(\ref{var1}) and (\ref{var2}) that we further find the inhomogeneous frequency shift $v_n(t)$ and intersite nonlinearity $g(t)$  in the form
 \bes \label{coe} \bee
 \label{coe2} &&
  g(t)=\rho_0^{-2}\exp\left[2\int^t_0\gamma(s)\,ds\right], \\
  \label{coe1}&&
v_n(t)=v_1(t)n+v_0(t),
\ene \ees
where we have introduced two functions in the inhomogeneous frequency shift $v_n(t)$ in the form
\bes \bee \label{v1}
&& v_1(t)=\frac{\alpha(t)\dot{\beta}(t)-\dot{\alpha}(t)\beta(t)}{2[\alpha^2(t)+\beta^2(t)]}, \\ \vspace{1in}
&& \label{v0} v_0(t)=a\left[\frac{\alpha^2(t)+\beta^2(t)}{a^2+b^2}\right]^{1/2}-\frac{\dot{p}_0(t)}{2},\ene\ees
where $p_0(t)$ is an arbitrary differentiable function of time.

It follows from Eqs.~(\ref{coe2})-(\ref{v0}) that, in these coefficients of Eq.~(\ref{ghirota}), the intersite nonlinearity $g(t)$ (the inhomogeneous frequency shift $v_n(t)$) is related to  the gain or loss term $\gamma(t)$ (the tunnel coupling $\Lambda(t)=\alpha(t)+i\beta(t)$). This means that only two varying coefficients (e.g., $\gamma(t)$ and
$\Lambda(t)=\alpha(t)+i\beta(t)$) are left free. Moreover, it follows from Eq.~(\ref{coe2}) that the intersite nonlinearity $g(t)$ is always positive (i.e., the attractive intersite nonlinearity). In addition, it follows from Eq.~(\ref{var1}) that the gain or loss term $\gamma(t)$ can also control the function $\rho(t)$, which is used to modulate the amplitude of the complex field $\Psi_n(t)$.

 For the inhomogeneous frequency shift $v_n(t)$ given by Eq.~(\ref{coe1}), when $\alpha(t)\not=c\beta(t)$ with $c$ being a constant, the inhomogeneous frequency shift $v_n(t)$ is a linear function of space $n$ with coefficients being functions of time. In the absence of the discrete space $n$ in the inhomogeneous frequency shift, i.e., $v_n(t)\equiv v(t)$, which means that $\dot{p}_1(t)=0$ on the basis of Eq.~(\ref{sys4}),
  there exist two cases to be discussed:

 i) If $p_1(t)\neq 0$ in which we have $p_1(t)=p_1={\rm const}\not=0$, then this means that $\varphi_n(t)$ is still a linear function of the discrete space $n$, i.e., $\varphi_n(t)=p_1n+p_0(t)$. In this case, the variable function $\tau(t)$ and the inhomogeneous frequency shift $v_n(t)$ are given by
 \bes \bee
\nonumber &&\tau(t)=\frac{\int^t_0\alpha(s)ds}{a\cos(p_1)+b\sin(p_1)}, \qquad\qquad \vspace{0.1in} \\
\nonumber && v_n(t)=\frac{a\alpha(t)}{a\cos(p_1)+b\sin(p_1)}-\frac{\dot{p}_0(t)}{2}, \ \ \
\ene \ees
 and $\rho(t),\, g(t)$ are same as Eqs.~(\ref{var1}) and (\ref{coe2});

 ii) If $p_1(t)=0$, i.e., $\tan^{-1}\{[b\alpha(t)-a\beta(t)]/[a\alpha(t)+b\beta(t)]\}=0$, which means that
the relation for the coefficients in Eq.~(\ref{ghirota}), $\alpha(t)=(a/b)\beta(t)$, is required and $\varphi_n(t)$ is only a functions of time, i.e.,
$\varphi_n(t)\equiv p_0(t)$, then the  variable function $\tau(t)$ and the inhomogeneous frequency shift $v_n(t)$ are given by the  form
 \bee
\nonumber  \tau(t)=\frac1a\int^t_0\alpha(s)ds, \quad v_n(t)=\alpha(t)-\frac{\dot{p}_0(t)}{2},
\ene
and $\rho(t),\, g(t)$ are same as Eqs.~(\ref{var1}) and (\ref{coe2}), where $\gamma(t),\ \alpha(t)$ and $p_0(t)$ are free functions of time, and $a,\ b,\ \rho_0$ are all free parameters.

\section{Nonautonomous discrete rogon solutions and interaction}

In general, we have a large degree of freedom in choosing the coefficients of similarity transformation (\ref{tran}) and Eq.~(\ref{ghirota}). As a consequence, we can obtain an infinitely large family of exact solutions of the generalized ALH lattice with variable coefficients given by Eq.~(\ref{ghirota}) in terms of exact solutions of the discrete Hirota equation (\ref{hirota}) and transformation (\ref{tran}). In particular, if we consider discrete rogon solutions of Eq.~(\ref{hirota}) as seeding solutions, then we can obtain many types of nonautonomous (including arbitrary time-dependent functions) discrete rogon (rogue wave) solutions of Eq.~(\ref{ghirota}). As two representative examples, we consider the lowest-order discrete rogon solutions of Eq.~(\ref{hirota}) as two examples~\cite{al} to study the dynamics of nonautonomous discrete rogon solutions of Eq.~(\ref{ghirota}).

\subsection{Nonautonomous discrete one-rogon solution}

Firstly, based on the similarity transformation (\ref{tran}) and one-rogon solution of the discrete Hirota equation (\ref{hirota}), we present the nonautonomous discrete one-rogon solution (also known as the first-order rational solution) of Eq.~(\ref{ghirota}) in the form
\bee
 \Psi^{(1)}_n(t)\!=\!\rho_0\sqrt{\mu}\exp{\left\{-\int^t_0\gamma(s)ds+i[\varphi_n(t)+\hat{\varphi}_n(\tau)]\right\}}
 \hspace{1cm}\nonumber \ene
\bee  \times \left[1\!-\!\frac{4(1+\mu)\left[1+4i\mu \sqrt{a^2+b^2}\,\tau(t)\right]}{1+4\mu n^2+16\mu^2(1+\mu)(a^2+b^2)\tau^2(t)}\right],
  \label{solu1}
    \ene
where the part phase $\hat{\varphi}_n(\tau)$ is defined by
\bee\label{phin}
 \hat{\varphi}_n(\tau)\!=\!2\tau(t)\!\!\left[(1\!+\!\mu)\sqrt{a^2+b^2}\!-\!a\right]\!\!
 -\!n\tan^{\!-1}\!\left(b/a\right),\ \
\ene
$\mu$ is a positive parameter, the variable $\tau(t)$ is given by Eq.~(\ref{var3}), and the phase $\varphi_n(t)=p_1(t)n+p_0(t)$ with $p_1(t)$  given by Eq.~(\ref{var2}) and $p_0(t)$ being an arbitrary differentiable function of time.

To illustrate the wave propagations of the obtained nonautonomous discrete one-rogon solution (\ref{solu1}),
 we can choose these free parameters in the form
\bee \label{gamm}
\begin{array}{l}
 \alpha(t)=c_1\sin(2t), \quad
 \beta(t)=c_2\cos(t), \vspace{0.1in}\cr
 \gamma(t)=\gamma_0\sin(t)\cos^2(t), \vspace{0.1in}\cr
  a=b=\mu=\rho_0=1, \end{array}
 \ene
where $\gamma_0,\, c_{1,2}$ are constants.

\begin{figure}
\begin{center}
\vspace{0.1in}
{\scalebox{0.42}[0.5]{\includegraphics{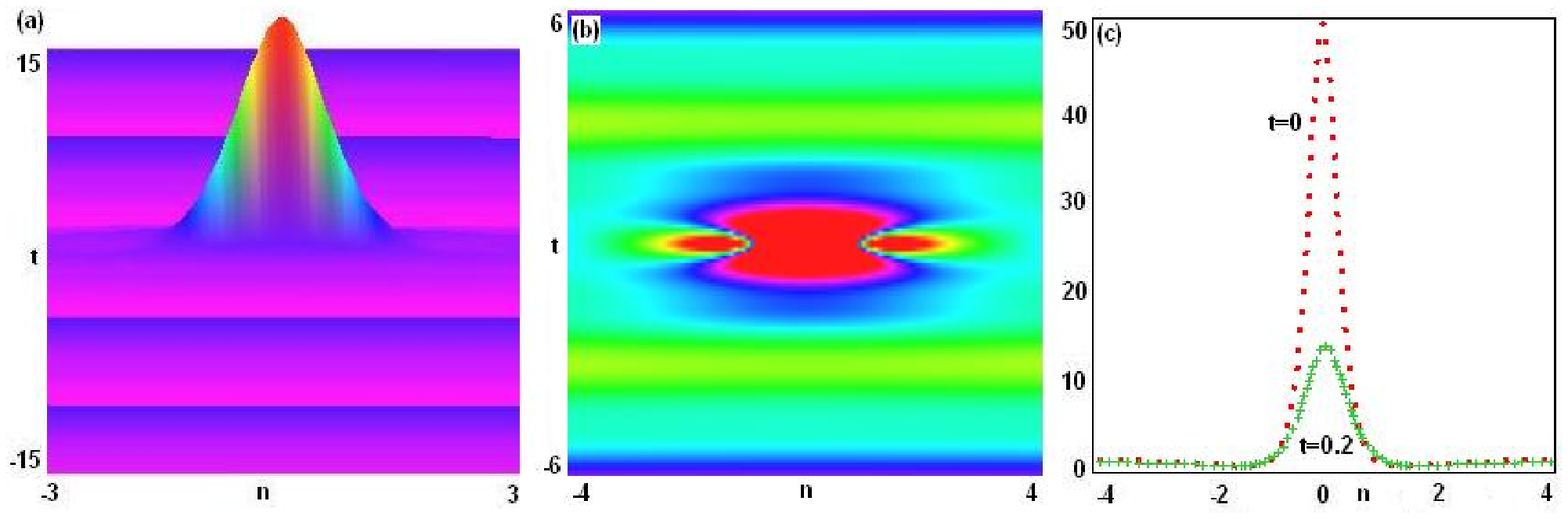}}}
\end{center}
\vspace{-0.15in} \caption{\small (color online).  Profiles of nonautonomous discrete one-rogon solution (\ref{solu1}) of the generalized ALH lattice with variable coefficients given by Eq.~(\ref{ghirota}) for the parameters given by Eq.~(\ref{gamm}) with $\gamma_0=c_1=c_2=1$. (a) the intensity distribution $|\Psi_n^{(1)}(t)|^2$ with ${\rm max}_{(n,t)}|\Psi_n^{(1)}(t)|^2\cong 4.7$, (b) the density distribution $|\Psi_n^{(1)}(t)|^2$, (c) the intensity distributions $|\Psi_n^{(1)}(t)|^2$ for $t=0,\, 0.2$, which means that the amplitude decreases as time increases, and the peak falls the lower position after time exceeds about $1$.} \label{fig:1}
\end{figure}

\begin{figure}
\begin{center}
\vspace{0.1in}
{\scalebox{0.42}[0.5]{\includegraphics{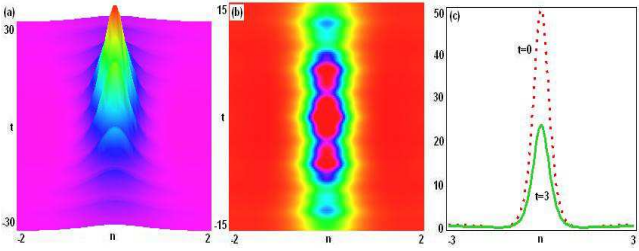}}}
\end{center}
\vspace{-0.15in} \caption{\small (color online).  Profiles of nonautonomous discrete one-rogon solution (\ref{solu1}) of the generalized ALH lattice with variable coefficients given by Eq.~(\ref{ghirota}) for the parameters given by Eq.~(\ref{gamm}) with $\gamma_0=1, \, c_1=0.01$, and $c_2=0.02$.  (a) the intensity distribution $|\Psi_n^{(1)}(t)|^2$ with ${\rm max}_{(n,t)}|\Psi_n^{(1)}(t)|^2\cong 43.2$, (b) the density distribution $|\Psi_n^{(1)}(t)|^2$, (c) the intensity distributions $|\Psi_n^{(1)}(t)|^2$ for $t=0,\, 3$, which means that the amplitude decreases as time increases, and the peak falls the lower position after time exceeds about $70$.} \label{fig:2}
\end{figure}

Figure~\ref{fig:coe} depicts the profiles of the coefficient $v_1(t)=[\alpha(t)\dot{\beta}(t)-\dot{\alpha}(t)\beta(t)]/\{2[\alpha^2(t)+\beta^2(t)]\}$ of the first degree term of the inhomogeneous frequency shift $v_n(t)$ given by Eq.~(\ref{v1}), the attractive intersite nonlinearity $g(t)$ given by Eq.~(\ref{coe2}), and  the gain or loss term $\gamma(t)$ vs time for the parameters given by Eq.~(\ref{gamm}). The evolution of the intensity distribution for the one-rogon solution given by Eq.~(\ref{solu1}) is illustrated in Fig.~\ref{fig:1} for parameters $\gamma_0=c_1=c_2=1$. The discrete rogue wave solution is localized both in space and in time, thus revealing the usual discrete `rogue wave' features. However if we fix the coefficient $\gamma_0=1$ of the gain or loss term and adjust the coefficients $c_1=0.01,\ c_2=0.02$ of the tunnel couplings $\alpha(t)$ and $\beta(t)$ given by Eq.~(\ref{gamm}), then the evolution of the intensity distribution for the one-rogon solution is changed (see Fig.~\ref{fig:2}), and it follows from Fig.~\ref{fig:2} that the discrete one-rogon solution in this case is localized in space and keep the localization longer in time than usual rogue waves (see \cite{al}). Moreover, it follows from Figs.~\ref{fig:1}(c) and \ref{fig:2}(c) that the amplitude of the discrete one-rogon solution decreases as time increases, and the amplitude in Fig.~\ref{fig:1}(c) is decreased faster than one in Fig.~\ref{fig:2}(c) as time increases.

\subsection{The interaction between nonautonomous discrete rogon solutions}

Here we consider the interaction between nonautonomous discrete rogon solutions. To do so, we apply a second-order rational solution of the discrete Hirota equation
(\ref{hirota}) to the similarity transformation (\ref{tran}) such that we can obtain the nonautonomous discrete
two-rogon solution of Eq.~(\ref{ghirota}) in the form
 \bee
 \Psi^{(2)}_n(t)=\rho_0\sqrt{\mu}\exp{\left\{\!-\!\int^t_0\!\gamma(s)ds+i[\varphi_n(t)+\hat{\varphi}_n(\tau)]\right\}} \quad \nonumber\\
  \times\!\!\left[1\!-\frac{\!12(1\!+\!\mu)\big[\mathcal{P}^{(2)}_n(\tau)\!+\!i\sqrt{\frac{\mathscr{T}(\tau)}{1+\mu}}
   \mathcal{Q}^{(2)}_n(\tau)\big]}{\mathcal{H}^{(2)}_n(\tau)}\right], \qquad
      \label{solu2} \ene
which displays the interaction between nonautonomous discrete rogon solutions, where $\mu$ is a positive parameter, the functions $\mathcal{P}^{(2)}_n(\tau),\, \mathcal{Q}^{(2)}_n(\tau)$ and $\mathcal{H}^{(2)}_n(\tau)$ are all polynomials of space and time given by
\bee \nonumber
 \begin{array}{l}
  \mathcal{P}^{(2)}_n(\tau)=5\mathscr{T}^2+6(\mathscr{N}+2\mu+3)\mathscr{T}+\mathscr{N}^2 \qquad\qquad
 \vspace{0.1in}\cr
  \qquad\qquad\quad  +(6-4\mu)\mathscr{N}-3(4\mu+1),
\end{array}
 \ene
 \vspace{-0.15in} \bee \nonumber
 \begin{array}{l}
 \mathcal{Q}^{(2)}_n(\tau)=\mathscr{T}^2+2(\mathscr{N}+1)\mathscr{T}+\mathscr{N}^2 \qquad\qquad\quad \qquad
\vspace{0.1in}\cr
  \qquad\qquad\quad  -(16\mu+6)\mathscr{N}-3(8\mu+5),
\end{array}
 \ene
\vspace{-0.15in} \bee \nonumber
 \begin{array}{l}
 \mathcal{H}^{(2)}_n(\tau)\!=\!\mathscr{T}^3\!+\!3(\mathscr{N}\!+\!8\mu+9)\mathscr{T}^2
 \!+\!3(\mathscr{N}^2\!-\!6\mathscr{N} \qquad
\vspace{0.1in}\cr
\qquad\qquad\quad \!-\!16\mu \mathscr{N}\!+\!48\mu^2\!+\!72\mu\!+\! 33)\mathscr{T}\!+\!\mathscr{N}^3
\vspace{0.1in}\cr
\qquad\qquad\quad  \!+\!(3-8\mu)\mathscr{N}^2\!+\!(27+24\mu+16\mu^2)\mathscr{N}\!+\!9,
 \end{array}
 \ene
where we have introduced $\mathscr{N}=4\mu n^2$ and $\mathscr{T}=16\mu^2(1+\mu)(a^2+b^2)\tau^2,$  $\tau\equiv \tau(t)$ is given by Eq.~(\ref{var3}), the part phase $\varphi_n(t)=p_1(t)n+p_0(t)$ with $p_1(t)$  given by Eq.~(\ref{var2}) and $p_0(t)$ being an arbitrary differentiable function of time, and the part phase $\hat{\varphi}_n(\tau)$ is given by Eq.~(\ref{phin}).

\begin{figure}[!ht]
\begin{center}
\vspace{0.1in}
{\scalebox{0.42}[0.5]{\includegraphics{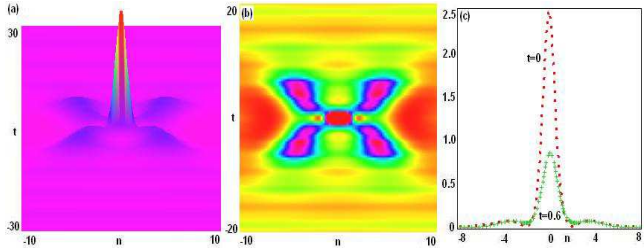}}}
\end{center}
\vspace{-0.2in} \caption{\small (color online).  Profiles of the interaction between nonautonomous discrete rogon solution given by Eq.~(\ref{solu2}) of the generalized ALH lattice with variable coefficients given by Eq.~(\ref{ghirota}) for the parameters given by Eq.~(\ref{gamm}) with $\gamma_0=1,\, c_1=2$, and $c_2=1$. (a) the intensity distribution $|\Psi_n^{(2)}(t)|^2$ with ${\rm max}_{(n,t)}|\Psi_n^{(2)}(t)|^2\cong 1.65$, (b) the density distribution $|\Psi_n^{(2)}(t)|^2$, (c) the intensity distributions $|\Psi_n^{(2)}(t)|^2$ for $t=0,\, 0.6$,  which means that the amplitude decreases as time increases, and the peak falls the lower position after time exceeds about $3$.} \label{fig:3}
\end{figure}

\begin{figure}[!ht]
\begin{center}
\vspace{0.05in}
{\scalebox{0.42}[0.5]{\includegraphics{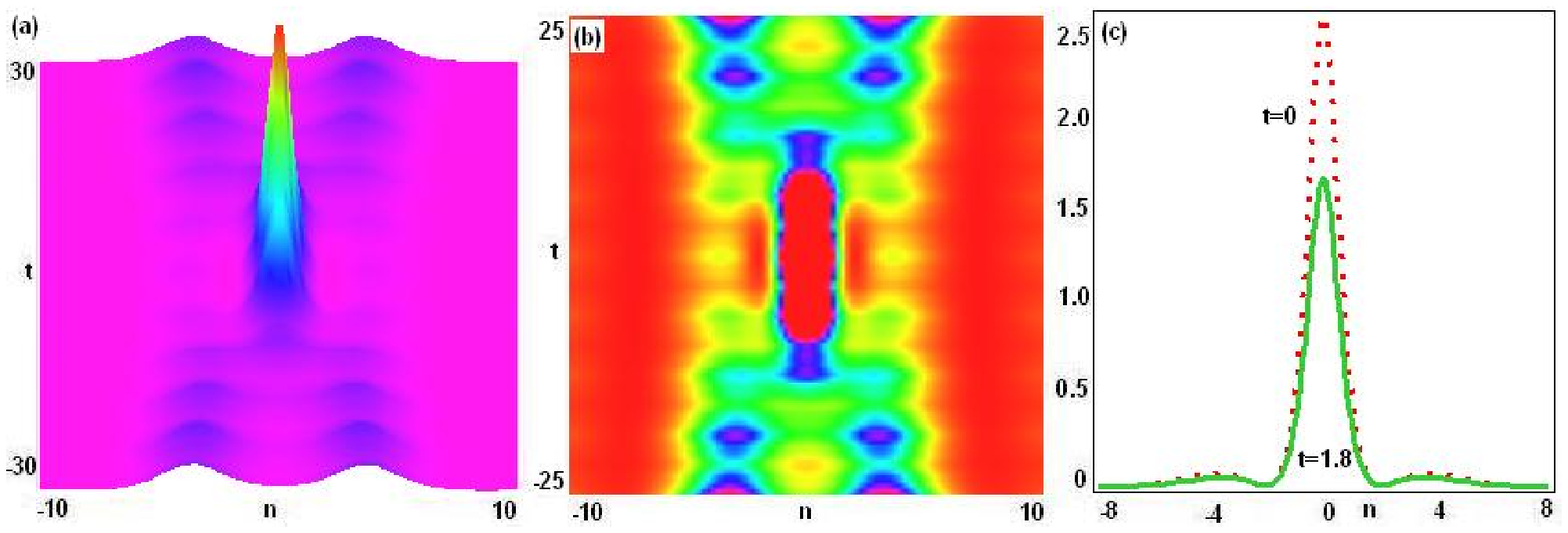}}}
\end{center}
\vspace{-0.2in} \caption{\small (color online).  Profiles of the interaction between nonautonomous discrete rogon solution given by Eq.~(\ref{solu2}) of the generalized ALH lattice with variable coefficients given by Eq.~(\ref{ghirota}) for the parameters given by Eq.~(\ref{gamm}) with  $\gamma_0=1,\, c_1=0.2$, and $c_2=0.1$. (a) the intensity distribution $|\Psi_n^{(2)}(t)|^2$ with ${\rm max}_{(n,t)}|\Psi_n^{(2)}(t)|^2\cong 2.25$, (b) the density distribution $|\Psi_n^{(2)}(t)|^2$, (c) the intensity distributions $|\Psi_n^{(2)}(t)|^2$ for $t=0,\, 1.8$,  which means that the amplitude decreases as time increases, and the peak falls the lower position after time exceeds about $65$.} \label{fig:4}
\end{figure}

Similarly, we can choose these free parameters given by Eq.~(\ref{gamm}) for the nonautonomous discrete two-rogon solution given by Eq.~(\ref{solu2}) except for $\mu=1/16$.
Figures~\ref{fig:3} and \ref{fig:4} depict the evolution of intensity distribution for the interaction between  nonautonomous discrete rogon solutions (discrete two-rogon solution) given by Eq.~(\ref{solu2}) for different parameters $\gamma_0,\, c_1$, and $c_2$, respectively. Moreover, we find that the interaction between nonautonomous discrete rogue waves is elastic.
Figure~\ref{fig:3} shows that the nonautonomous discrete two-rogon solution is localized both in space and in time, thus revealing the usual discrete `rogue wave' features for the chosen parameters $\gamma_0=1,\, c_1=2,\,c_2=1$, but
if we fix the coefficient $\gamma_0=1$ of the gain or loss term and adjust the coefficients $c_1=0.2,\ c_2=0.1$ of the tunnel couplings $\alpha(t)$ and $\beta(t)$ given by Eq.~(\ref{gamm}),  then the evolution of the intensity distribution for the discrete two-rogon solution is changed (see Fig.~\ref{fig:4}), and it follows from
Fig.~\ref{fig:4} that the nonautonomous discrete two-rogon solution is localized in space and keep the localization longer in time than usual rogue waves.
Moreover, it follows from Figs.~\ref{fig:3}(c) and \ref{fig:4}(c) that the amplitude of the discrete two-rogon solution decreases as time increases, and the amplitude in Fig.~\ref{fig:3}(c) decreases faster than one in Fig.~\ref{fig:4}(c) as time increases.

\section{Conclusions}

In conclusion, we have studied nonautonomous discrete rogon solutions and their interaction in the generalized
Ablowitz-Ladik-Hirota lattice with the varying tunnel coupling, intersite nonlinearity, inhomogeneous frequency shift, and gain or loss term given by Eq.~(\ref{ghirota}) on the basis of the differential-difference similarity reduction (\ref{tran}). We found its some nonautonomous discrete rogon solutions when the intersite nonlinearity $g(t)$ (the inhomogeneous frequency shift $v_n(t)$) is related to  the gain or loss term $\gamma(t)$ (the tunnel coupling $\Lambda(t)=\alpha(t)+i\beta(t)$) (see Eqs.~(\ref{coe2}) and (\ref{coe1})). This denotes that only two coefficients (i.e., $\gamma(t)$ and $\Lambda(t)=\alpha(t)+i\beta(t)$) are left free.

In particular, we have studied wave propagations of nonautonomous discrete rogon solutions and interaction for some chosen parameters, which exhibit complicated rogue wave construes. For the given periodic gain or loss term, when the amplitude of the tunnel coupling coefficient between sites decreases, these nonautonomous discrete rogon solutions are localized in space and keep the localization longer in time, which differ from the usual discrete rogue waves of nonlinear discrete equations (e.g., the AL lattice and the discrete Hirota equation)~\cite{al}.

Moreover, nonautonomous discrete rogon solutions and interaction may provide more documents to further understand the physical mechanism of discrete rogue wave phenomena. The approach may also be extended to other discrete nonlinear lattices with variable coefficients for studying their discrete rogue wave solutions and wave propagations.

\acknowledgments

 This work was supported by the  NSFC under Grant Nos. 60821002F02, 11071242, 10874235, 10934010,
60978019, the NKBRSFC under Grants Nos. 2009CB930701,
2010CB922904 and 2011CB921500.



\end{document}